\documentclass[runningheads]{llncs}
\usepackage[T1]{fontenc}
\usepackage{graphicx}
\usepackage{booktabs}
\usepackage{tikz}
\usetikzlibrary{arrows.meta,positioning,fit,calc}
\usepackage{array}
\usepackage{tabularx}
\usepackage{multirow}
\usepackage{placeins}
\usepackage{url}
\begin{document}
\title{Towards End-to-End Privacy-Preserving AI for Real-Time ECG Monitoring}
\titlerunning{Privacy-Preserving AI for Real-Time ECG Monitoring}
\author{Beyaz{\i}t Bestami Y\"uksel\inst{1}}
\authorrunning{B.\ B.\ Y\"uksel}
\institute{Department of Computer Engineering,\\
Istanbul Technical University,\\
Istanbul, T\"urkiye\\
\email{yukselbe18@itu.edu.tr}}
\maketitle
\noindent\textbf{PhD status.}
PhD graduate, February 2026.
The dissertation was defended on 17 February 2026.
\begin{abstract}
Healthcare artificial intelligence is shifting from centralized archives toward distributed
pipelines that combine wearables, edge processing, cloud analytics, and multi-institutional
learning.
In this regime, electrocardiogram (ECG) streams cross multiple trust boundaries, so privacy
cannot be reduced to a single transmission control.
Existing privacy-enhancing technologies typically secure isolated stages and leave gaps
between acquisition, analysis, and institutional collaboration.

This doctoral research develops a unified lifecycle framework for end-to-end
privacy-preserving ECG AI.
Rather than introducing another privacy mechanism, it orchestrates complementary
privacy-enhancing technologies across the clinical ECG lifecycle, allocating each mechanism
to the trust boundary where its security, computational, and operational properties are most
appropriate.
Three interoperable modules instantiate the framework: AES with CKKS-based homomorphic
statistics for secure monitoring; adaptive chaotic protection for low-latency edge
transmission; and federated learning with homomorphic aggregation under heterogeneous
partitions.
Across prototype monitoring, signal-security analysis, and simulated collaboration, the
programme shows that privacy layers can be composed without collapsing diagnostic utility,
while exposing overhead and fidelity limits that constrain clinical translation.
\end{abstract}

\section{Introduction and Motivation}
\label{sec:introduction}

Healthcare AI is moving beyond offline analysis of hospital archives.
Remote monitoring, wearables, and edge--cloud orchestration distribute sensing, inference,
and model improvement across devices, networks, and institutions.
This improves coverage, but multiplies exposure points for sensitive physiological data and
for models derived from them.

ECG monitoring exemplifies the shift.
Continuous streams support rhythm assessment and remote follow-up, yet become
privacy-critical once coupled with cloud storage, remote visualization, and cross-site
learning.
Channel encryption alone is then insufficient: threats arise during acquisition and
transmission, during stored-data analytics, and during exchange of model updates among
institutions with heterogeneous schemas.

In distributed AI pipelines, raw signals, intermediate representations, model parameters,
and learned updates can all become privacy-sensitive assets.
Privacy therefore becomes a system-level property rather than a single control.

This dissertation addresses that systems problem~\cite{yuksel2026thesis}.
Three published research lines provide the technical building blocks: ECG-PPS for
acquisition, monitoring, AES-protected storage and transmission, and homomorphic
statistics~\cite{yuksel2024ecgpps}; HEART for adaptive chaotic protection with a learnable
key generator~\cite{yuksel2026heart,yuksel2024advancing}; and FL-HE for federated ECG learning with
homomorphic update protection under horizontal, vertical, and combined
partitions~\cite{yuksel2026flhe}.
Evidence is synthesized conservatively, distinguishing prototype evaluation, retrospective
validation, simulated collaboration, and expert review from prospective clinical trials.

\subsection{Main Dissertation Contributions}
\label{sec:main-contributions}
\vspace{-0.35em}
\begin{itemize}\setlength{\itemsep}{0.15em}\setlength{\parsep}{0pt}\setlength{\topsep}{0.2em}\setlength{\partopsep}{0pt}
\item A trust-boundary allocation within a unified privacy framework that decomposes
continuous ECG AI into distinct trust boundaries rather than treating privacy as a single
end-to-end control.
\item A lifecycle-oriented allocation of PETs that matches each mechanism to the threat
surface and operational constraints of its intended stage.
\item Experimental validation of three interoperable privacy modules (ECG-PPS, HEART,
and FL-HE) as complementary instantiations of the same framework.
\item Identification of privacy--latency--utility trade-offs that constrain uniform
application of heavy cryptographic protection under continuous monitoring.
\end{itemize}

\section{Research Gap and Doctoral Questions}
\label{sec:research-gap}

The preceding motivation raises a concrete research gap.
Privacy-preserving biomedical AI has advanced along largely separate tracks: conventional
encryption for transport and storage; chaos-based signal protection; fully homomorphic
encryption (FHE) for computation on ciphertexts; and federated learning (FL) without
raw-data pooling.
Each track is valuable, yet previous work typically optimizes one PET independently at a
single boundary.
AES-style protection still exposes plaintext when analytics require decryption; chaotic
schemes can be low-latency but are not substitutes for standard cryptography; homomorphic
methods enable encrypted statistics or protected updates at nontrivial cost; and FL reduces
data sharing but remains exposed to update-level inference without further defences.
Securing one hop therefore leaves residual exposure elsewhere in continuous ECG workflows.

The unresolved scientific problem is not whether PETs work individually, but how
heterogeneous PETs should be orchestrated across the complete clinical AI lifecycle.
Prior studies rarely examine this allocation under joint privacy, latency, and
diagnostic-utility constraints.
The doctoral gap is therefore how complementary PETs can be composed across distinct trust
boundaries of an end-to-end ECG pipeline so that privacy remains manageable without
collapsing clinical feasibility.

\subsection{Design Principles of the Doctoral Framework}
\label{sec:design-principles}

Guided by that gap, six principles structure the programme:
(i)~lifecycle-aware privacy across the full ECG pathway;
(ii)~boundary-specific protection matched to local threat surfaces and budgets;
(iii)~real-time feasibility under measured monitoring envelopes;
(iv)~composable privacy in which AES, adaptive chaotic protection, FHE/CKKS, and FL address
distinct assets rather than rival one another;
(v)~a translational objective with explicit threat assumptions and honest overhead reporting;
and (vi)~evidence-aware translation that distinguishes prototype, retrospective, simulated,
and expert-review evidence from prospective clinical readiness.

\subsection{Doctoral Research Questions}
\label{sec:research-questions}
\begin{enumerate}\setlength{\itemsep}{0.15em}\setlength{\parsep}{0pt}\setlength{\topsep}{0.25em}
\item How should privacy responsibilities be decomposed across the ECG lifecycle so that
acquisition and transmission, secure analytics, and multi-institutional learning are coupled
stages of one system?
\item Which combinations of conventional encryption, adaptive signal protection, and
homomorphic computation remain compatible with continuous monitoring while preserving
essential diagnostic statistics and workflow usability?
\item How can collaborative ECG learning proceed under horizontal, vertical, and combined
partitions when updates must be protected against honest-but-curious aggregation?
\item Where do overhead, fidelity, heterogeneity, and validation scope constrain translation
from prototype pipelines to prospective clinical deployment?
\end{enumerate}

\paragraph{Doctoral contribution beyond the individual publications.}
The answers to these questions do not reduce to the three published methods themselves.
ECG-PPS, HEART, and FL-HE each address a different technical problem and could, in
principle, have been reported independently.
The systems-oriented contribution is their conceptual integration into one coherent
framework for continuous ECG AI~\cite{yuksel2026thesis}.
That integration yields a design approach for assigning complementary PETs to distinct trust
boundaries according to local security, computational, and operational requirements.
Without this organizing rationale, the publications remain adjacent technical advances
rather than a unified doctoral programme.
The originality of this dissertation lies not in proposing a single superior privacy
mechanism, but in establishing a systems engineering perspective for orchestrating
heterogeneous privacy-enhancing technologies according to the security, computational, and
operational requirements of the complete clinical ECG lifecycle.
The following section shows how this perspective is realized through the unified lifecycle
framework.

\section{A Unified Lifecycle Framework for Privacy-Preserving ECG AI}
\label{sec:unified-framework}

\begin{figure}[t]
\centering
\includegraphics[width=\linewidth]{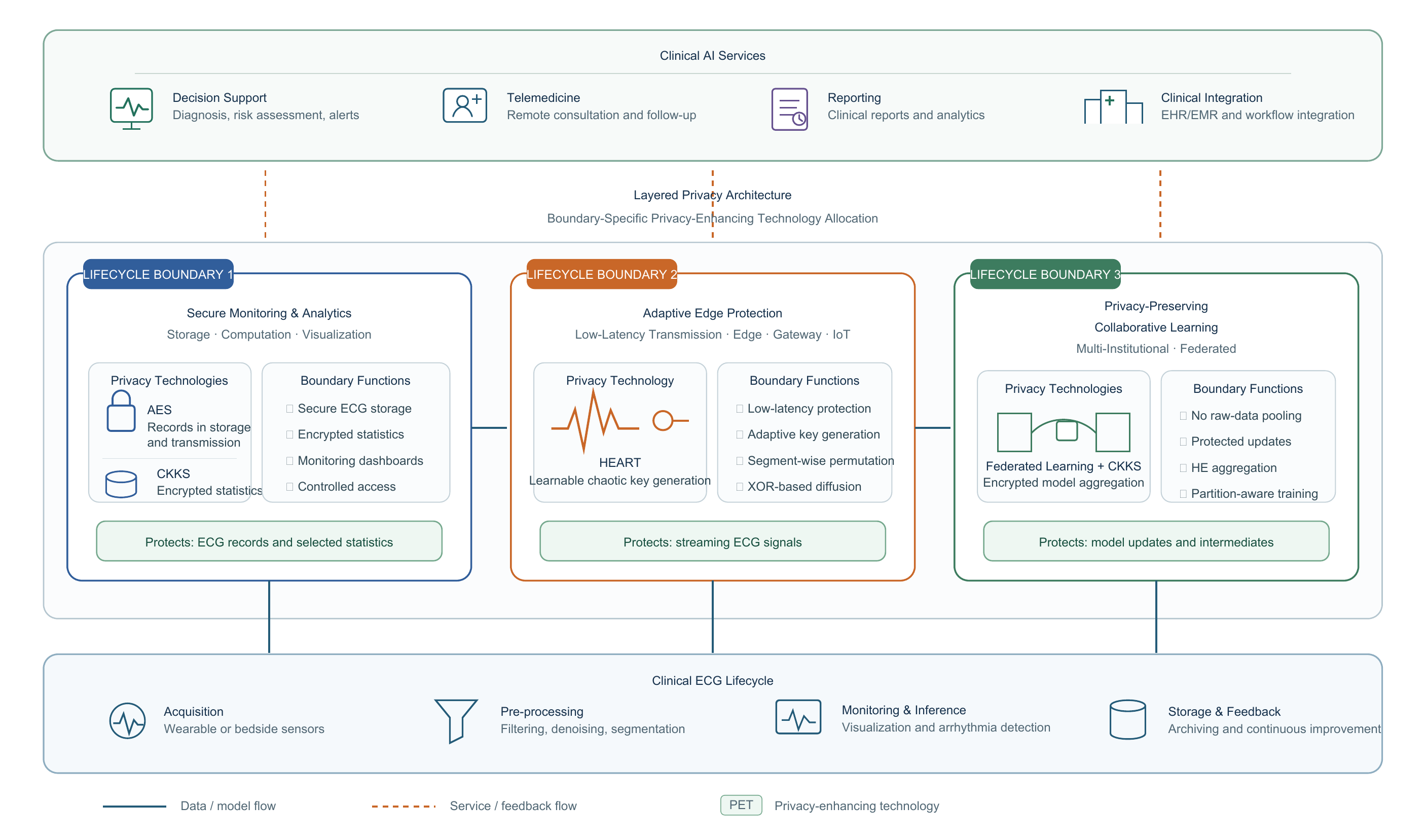}
\caption{Layered privacy architecture across the clinical ECG lifecycle.
ECG-PPS provides AES-protected monitoring with CKKS-based encrypted
statistics~\cite{yuksel2024ecgpps}; HEART adds adaptive chaotic edge
protection~\cite{yuksel2026heart}; and FL-HE enables collaborative learning with
CKKS-protected updates~\cite{yuksel2026flhe}.
The formal B1--B3 security-boundary allocation is detailed later
(Fig.~\ref{fig:boundary-mapping})~\cite{yuksel2026thesis}.}
\label{fig:lifecycle}
\end{figure}

Building on that claim, the dissertation is not searching for one universally optimal
privacy mechanism.
Instead, it introduces a design approach for selecting and composing complementary PETs
according to lifecycle requirements and maps those PETs onto the unified framework for
continuous ECG AI (Fig.~\ref{fig:lifecycle}).
Each layer is motivated by a distinct protected asset and threat surface; layers compose
rather than compete.
Because stages impose different threat models, latency constraints, and operational budgets,
heterogeneous PET composition matters more than maximizing any single method in isolation.
ECG-PPS, HEART, and FL-HE are therefore treated as interoperable modules of one
trust-boundary allocation within that framework, not as alternative solutions to the same
problem.

\paragraph{Boundary allocation.}
HEART supplies millisecond-scale adaptive chaotic protection at the source
edge~\cite{yuksel2026heart}; it complements, rather than replaces, standard cryptography and
is evaluated empirically rather than via formal IND-CPA proofs.
ECG-PPS combines AES-protected storage and transmission with CKKS-based FHE for
statistical operations~\cite{yuksel2024ecgpps}; disease detection uses retrospectively
trained MIT-BIH models, and reported FHE evidence concerns statistical agreement rather than
full neural inference on ciphertexts.
FL-HE trains on PTB-XL under simulated horizontal, vertical, and combined partitions, with
CKKS protecting updates under honest-but-curious assumptions~\cite{yuksel2026flhe}.
The programme-level claim is an allocation strategy across trust boundaries: adaptive
source protection where latency dominates, conventional cryptography for communication and
storage, and homomorphic FL where institutions collaborate without sharing raw
records~\cite{yuksel2026thesis}.
Unlike the individual publications, which each optimize one privacy mechanism, the
dissertation contributes the systems-oriented rationale for when, where, and why each PET
should be deployed in an end-to-end clinical AI pipeline.

\section{Research Modules Instantiating the Framework}
\label{sec:contributions}

The three published modules are summarized next as interoperable components of one doctoral
programme.

\subsection{ECG-PPS: Monitoring Workflow with AES and Homomorphic Statistics}
\label{sec:ecg-pps}
\paragraph{Motivation.}
Continuous monitoring needs a pathway that acquires signals, supports rhythm-oriented
detection, stores data for remote review, and permits limited analytics without treating
cloud plaintext as inevitable.

\paragraph{Contribution.}
ECG-PPS implements three-lead serial acquisition with real-time visualization, a CNN
arrhythmia detector trained on MIT-BIH, AES-protected storage and transmission (including an
ECDH+AES/TLS variant), and CKKS-based FHE for statistical operations on encrypted
data~\cite{yuksel2024ecgpps}.
Protected assets are ECG records in transit or storage and selected encrypted statistics.

\paragraph{Evidence.}
Thesis synthesis reports encryption and transmission delays below 50\,ms per
segment~\cite{yuksel2026thesis}.
Encrypted-versus-plaintext agreement includes R-peak 100\%, mean 99.17\%, SD 98.11\%,
median 100\% (extrema: min 93.75\%, max 94.12\%)~\cite{yuksel2024ecgpps}.

\paragraph{Limits.}
FHE cost, MIT-BIH dependence, three-lead scope, and network availability constrain
translation; validation is prototype or retrospective, not prospective.
Secure monitoring therefore cannot be reduced to communication security alone.

\paragraph{Framework role.}
Implements the secure monitoring and analytics boundary of the unified framework.

\clearpage
\subsection{HEART: Adaptive Chaotic Protection at the Edge}
\label{sec:heart}
\paragraph{Motivation.}
Edge streaming needs confidentiality compatible with continuous acquisition.

\paragraph{Contribution.}
HEART estimates logistic-map parameters $(r,x_0)$ from ECG-segment statistics (optional ML
refinement under noise) and applies permutation with XOR diffusion~\cite{yuksel2026heart}.
Security arguments are empirical (entropy, correlation, NIST frequency tests, sensitivity,
noise and occlusion), not formal reductions.

\paragraph{Evidence.}
Entropy $7.6$--$7.8$\,bits; MSE~$\approx5\times10^{-6}$, PSNR~$>52$\,dB, MAE~$\approx0.002$;
encryption and decryption $\approx1.2$\,ms and $0.1$\,ms; end-to-end $\approx336$--$339$\,ms
per 300-sample segment (acquisition-dominated)~\cite{yuksel2026heart,yuksel2026thesis}.

\paragraph{Limits.}
Global fidelity does not imply preservation of all clinical morphology markers after
decryption; chaotic protection is not a drop-in substitute for standard cryptography, and
empirical evaluation must remain distinct from formal cryptographic guarantees.

\paragraph{Framework role.}
Implements adaptive source-boundary protection for low-latency ECG transmission.

\subsection{FL-HE: Multi-Institutional Learning with Protected Updates}
\label{sec:fl-he}
\paragraph{Motivation.}
Institutions often cannot share raw ECG repositories.
FL reduces raw-data movement but exposes update leakage under honest-but-curious
aggregation.

\begin{figure}[t]
\centering
\includegraphics[width=\linewidth]{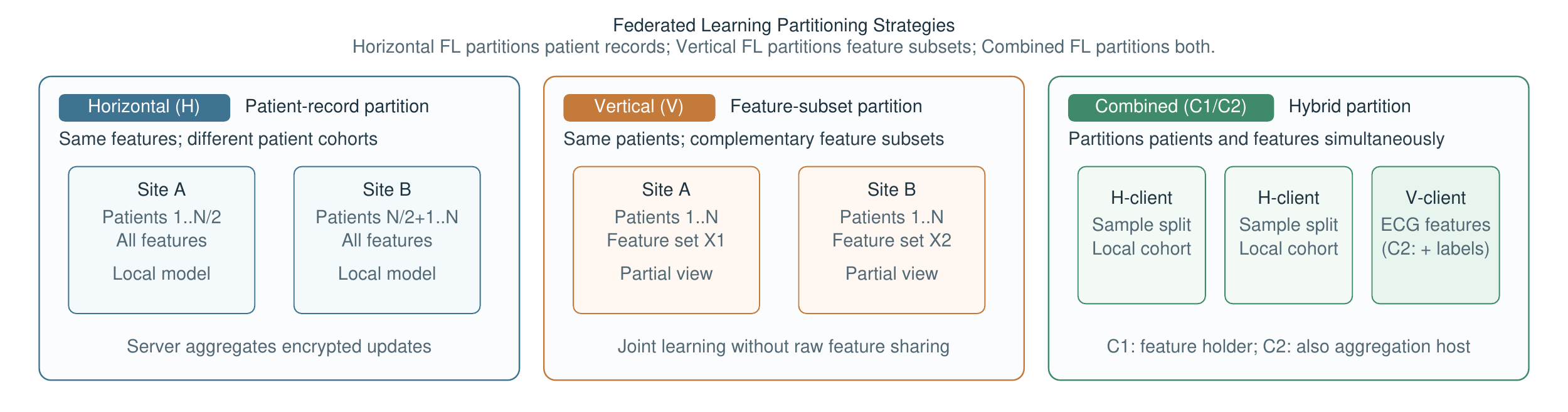}
\caption{Federated learning partitioning strategies evaluated in this work.
Horizontal FL partitions patient records, Vertical FL partitions feature subsets, and
Combined FL partitions both simultaneously.}
\label{fig:fl-partitions}
\end{figure}

\paragraph{Contribution.}
FL-HE trains on PTB-XL ($18{,}869$ patients; $21{,}799$ recordings) under simulated
multi-institutional partitions, using CKKS on updates and intermediates
~\cite{yuksel2026flhe,yuksel2026thesis}.
Four collaboration settings are evaluated (Fig.~\ref{fig:fl-partitions}):
\textbf{H}: same features, different patient cohorts;
\textbf{V}: same patients, complementary feature subsets;
\textbf{C1}: horizontal sample split plus a vertical client holding ECG features for all
patients;
\textbf{C2}: hybrid partitioning where the vertical client also acts as the aggregation
server and holds outcome labels.

\begin{figure}[t]
\centering
\includegraphics[width=\linewidth]{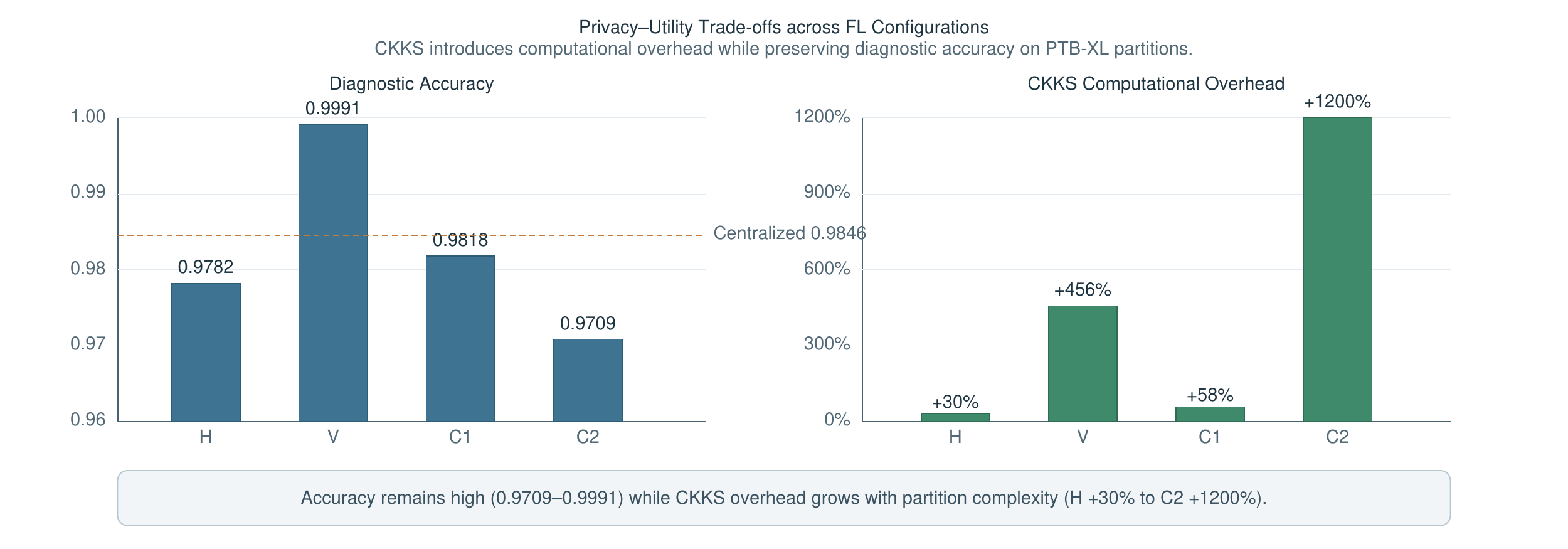}
\caption{Summary of the privacy--utility trade-offs observed across different federated
learning configurations.
CKKS introduces computational overhead while preserving diagnostic accuracy.}
\label{fig:tradeoff}
\end{figure}

\paragraph{Evidence.}
A custom backbone reports centralized accuracy $0.9846$ (F1 $0.9879$).
Under CKKS-protected FL, accuracies are $0.9782$ (H), $0.9991$ (V), $0.9818$ (C1), and
$0.9709$ (C2), with CKKS overheads $+30\%$, $+456\%$, $+58\%$, and $+1200\%$
(Table~\ref{tab:fl-scenarios}; Fig.~\ref{fig:tradeoff})~\cite{yuksel2026flhe}.
\begin{table}[t]
\caption{FL-HE accuracy and CKKS overhead across partition scenarios
~\cite{yuksel2026flhe}.}
\label{tab:fl-scenarios}
\centering
\small
\setlength{\tabcolsep}{12pt}
\begin{tabular*}{\linewidth}{@{\extracolsep{\fill}}llcc@{}}
\toprule
\textbf{Scenario} & \textbf{Partition type} & \textbf{Accuracy} & \textbf{CKKS overhead} \\
\midrule
H  & Horizontal   & 0.9782 & $+30\%$ \\
V  & Vertical     & 0.9991 & $+456\%$ \\
C1 & Combined-1   & 0.9818 & $+58\%$ \\
C2 & Combined-2   & 0.9709 & $+1200\%$ \\
\bottomrule
\end{tabular*}
\end{table}
Cardiologist review addressed dataset usage and diagnostic behaviour (expert review, not a
prospective multi-centre trial).

\paragraph{Limits.}
Partitions simulate heterogeneity; vertical and complex combined HE costs, especially C2,
limit strict real-time training loops.
Feasibility therefore depends on partition topology, overhead, and institutional workflow
assumptions as well as update protection.

\paragraph{Framework role.}
Implements the privacy-preserving collaborative-learning boundary through encrypted model
aggregation.

\section{Experimental Evidence and Cross-Module Analysis}
\label{sec:evidence}

The modules are next examined jointly to assess whether trust-boundary allocation remains
feasible under measured privacy, latency, and utility constraints.
\begin{table}[t]
\caption{Cross-module comparison.}
\label{tab:cross-module}
\centering
\small
\setlength{\tabcolsep}{4pt}
\resizebox{\linewidth}{!}{%
\begin{tabular}{@{}lllllll@{}}
\toprule
\textbf{Module} & \textbf{Asset} & \textbf{Threat} & \textbf{Technique} & \textbf{Where} & \textbf{Evidence} & \textbf{Limit} \\
\midrule
ECG-PPS &
Records; stats &
Interception; plaintext analytics &
AES (+TLS variant); CKKS stats &
Edge+cloud &
R-peak 100\%, mean 99.17\%; $<$50\,ms &
FHE cost; MIT-BIH; 3-lead \\
\midrule
HEART &
Stream confidentiality &
Low-latency + static keys &
Adaptive chaos (perm.+XOR); LKG &
Edge &
Entropy 7.6--7.8; PSNR~$>$52\,dB; $\sim$1.2\,ms &
No formal proof; morphology drift \\
\midrule
FL-HE &
Updates; data locality &
Honest-but-curious agg.; silos &
FL + CKKS updates &
Clients+server &
Acc.\ 0.9709--0.9991; expert review &
Simulated sites; HE up to $1200\%$ \\
\bottomrule
\end{tabular}%
}
\end{table}

\noindent
Verified anchors include ECG-PPS statistical agreement (R-peak 100\%; mean 99.17\%;
SD~98.11\%; $<$50\,ms/seg.), HEART security/timing (entropy 7.6--7.8;
MSE~$\approx5\!\times\!10^{-6}$; E2E~$\approx$336--339\,ms), and FL-HE accuracy/overhead
(H/V/C1/C2: 0.9782/0.9991/0.9818/0.9709; $+30$/$+456$/$+58$/$+1200\%$).
Table~\ref{tab:cross-module} supports composability rather than dominance of one PET.
Caveats remain: high PSNR need not preserve all morphology markers; encrypted statistical
agreement need not imply ciphertext neural diagnosis; and simulated partition accuracies
need not imply live multi-hospital deployment.
These findings motivate the formal security-responsibility mapping summarized next.

\begin{figure}[t]
\centering
\includegraphics[width=\linewidth]{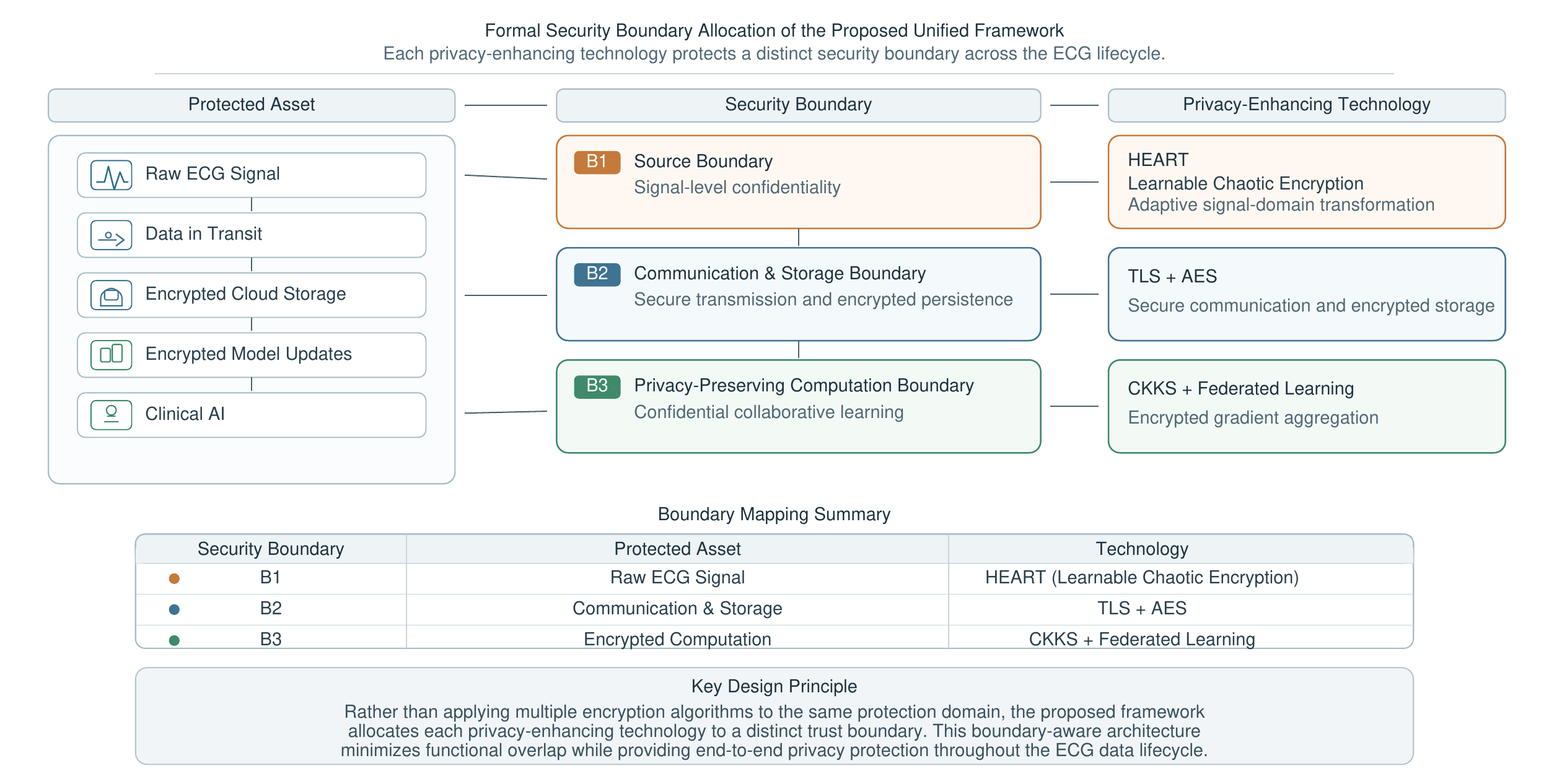}
\caption{Formal security boundary allocation of the proposed unified framework.
Each privacy-enhancing technology protects a distinct trust boundary across the ECG
lifecycle, minimizing functional overlap between complementary privacy-enhancing
technologies while enabling end-to-end privacy preservation.}
\label{fig:boundary-mapping}
\end{figure}
\noindent
Figure~\ref{fig:boundary-mapping} represents a formal allocation of security
responsibilities across trust boundaries, not another system-architecture diagram.
Learnable chaotic encryption is assigned to the source boundary; TLS and AES secure
communication and encrypted storage; and CKKS supports privacy-preserving federated
computation.
By binding each PET to a distinct responsibility, the allocation minimizes functional
overlap between mechanisms while still covering the ECG lifecycle end to end.
The technologies therefore complement rather than replace one another within a coherent
privacy architecture~\cite{yuksel2026thesis}.

\section{Open Challenges and Future Research}
\label{sec:challenges}

Even under this allocation, open issues follow the dissertation limits~\cite{yuksel2026thesis}.
End-to-end composition still needs stronger key management and latency accounting under
uncontrolled networks; chaotic threat models remain empirical; and HE/FL arguments centre on
honest-but-curious update exposure.
CKKS overheads from tens of percent to order-of-magnitude costs show that homomorphic
protection cannot be applied uniformly under binding monitoring constraints
(Fig.~\ref{fig:tradeoff}).
Simulated partitions and public datasets only approximate institutional heterogeneity;
regulatory alignment is a design motivation, not a certification claim.
Future work should examine adaptive privacy orchestration that selects PETs according to
computation, trust assumptions, network conditions, partitioning, and clinical workflow.

\section{Dissertation Status and Long-Term Goals}
\label{sec:status}
\enlargethispage{8\baselineskip}

\textbf{Applicant.} Beyaz{\i}t Bestami Y\"uksel, Department of Computer Engineering,
Istanbul Technical University, Istanbul, T\"urkiye.
\textbf{Status.} PhD graduate, February 2026; defended 17 February 2026.
\textbf{Title.} \textit{A Unified Framework for Secure and Intelligent ECG Signal Processing
via Chaotic Encryption, Fully Homomorphic Computation, and Federated
Learning}~\cite{yuksel2026thesis}.
Associated outputs~\cite{yuksel2024ecgpps,yuksel2026heart,yuksel2026flhe} realize the
lifecycle-oriented programme summarized above.
Next steps include tighter lifecycle integration, HE acceleration, morphology-preserving
refinements, and partition-realistic multi-site evaluation.

\section{Conclusion}
\label{sec:conclusion}

This dissertation argues that privacy-preserving medical AI should be designed from a
systems engineering perspective rather than by searching for one universally optimal privacy
mechanism.
By allocating adaptive chaotic edge protection, AES/TLS for communication and storage, and
federated learning with encrypted updates to distinct trust boundaries, it establishes a
framework-level approach to PET orchestration across the clinical ECG lifecycle.
Evidence supports feasibility under prototype and simulated settings, while measured
overhead and fidelity limits clarify the remaining distance to prospective
deployment~\cite{yuksel2026thesis}.

\FloatBarrier
\begin{credits}
\enlargethispage{2\baselineskip}
\subsubsection{\ackname}
All code and datasets associated with this work are available at
\url{https://github.com/bbyuksel}.

\subsubsection{\discintname}
The author has no competing interests to declare that are relevant to the content of this article.
\end{credits}
\bibliographystyle{splncs04}
\bibliography{references}
\end{document}